\documentclass{svproc}
\usepackage{url}
\usepackage[colorlinks=true,urlcolor=blue]{hyperref}
\newcommand*{\doi}[1]{\href{http://dx.doi.org/#1}{doi: #1}}

\usepackage{bm}
\usepackage{amssymb,amsmath,mathtools}
\usepackage[table,xcdraw]{xcolor}
\usepackage{graphicx}
\usepackage{subcaption}
\usepackage{algorithmic, algorithm}

\begin{document}
\mainmatter              
\title{Maximum Likelihood Estimation for a Markov-Modulated Jump-Diffusion Model}
\titlerunning{MLE for a Markov-Modulated Jump-Diffusion Model}  
%
\author{
Bor Reynoso\inst{1} 
\and F. Baltazar-Larios\inst{1}
\and Laura Eslava\inst{2}
}
\authorrunning{Bor Reynoso et al.} 
%

%
\institute{FC, UNAM,\\
\email{borreynoso@ciencias.unam.mx}
\and
IIMAS, UNAM}

\maketitle              

\begin{abstract}
We propose a method for obtaining maximum likelihood estimates (MLEs) of a Markov-Modulated Jump-Diffusion Model (MMJDM) when the data is a discrete time sample of the diffusion process, the jumps follow a Laplace distribution, and the parameters of the diffusion are controlled by a Markov Jump Process (MJP). The data can be viewed as incomplete observation of a model with a tractable likelihood function. Therefore we use the EM-algorithm to obtain MLEs of the parameters. We validate our method with simulated data.

The motivation for obtaining estimates of this model is that stock prices have distinct drift and volatility at distinct periods of time. The assumption is that these phases are modulated by macroeconomic environments whose changes are given by discontinuities or jumps in prices. This model improves on the stock prices representation of classical models such as the model of Black and Scholes or Merton's Jump-Diffusion Model (JDM). We fit the model to the stock prices of Amazon and Netflix during a 15-years period and use our method to estimate the MLEs.

\end{abstract}
\section{Introduction}

Modelling the stock market is a central topic of mathematical finance due to the importance of predicting values to help investors identify investments with better returns. It is generally assumed that the stock prices may be represented by a continuous-time stochastic process $\{S_t\}_{t\ge 0}$ and that we may observe its values at discrete times.  From these historical data, we may estimate uncertainties and predict stock market trends to help investors when making decisions.

One of the earliest model for $\{S_t\}_{t\ge 0}$ is the Geometric Brownian Motion (GBM) introduced by Black and Scholes in 1973 \cite{Merton73}. This process is defined by its initial value, drift and volatility; see equation \eqref{EGBM} for a detailed definition. 

The GBM is central to describe the behavior of a stock price in option pricing models. However, since the works in \cite{Cla73}, \cite{Man63}, and \cite{Man67}, it is recognized that the market dynamics can not be modelled by GBM with constant parameters of drift and volatility. One of its main drawbacks is that large and abrupt changes in the prices may not be accounted for without assuming that the process has an artificially large volatility. The source of these abrupt changes may be macroeconomic events such as changes in the rate of economic growth, gross domestic product, unemployment rates, and inflation. 

\begin{figure}
\centering
\subcaptionbox{Netflix Stock Price (USD) from July 1st 2005 to June 30 2020.}{\includegraphics[width=0.9\textwidth]{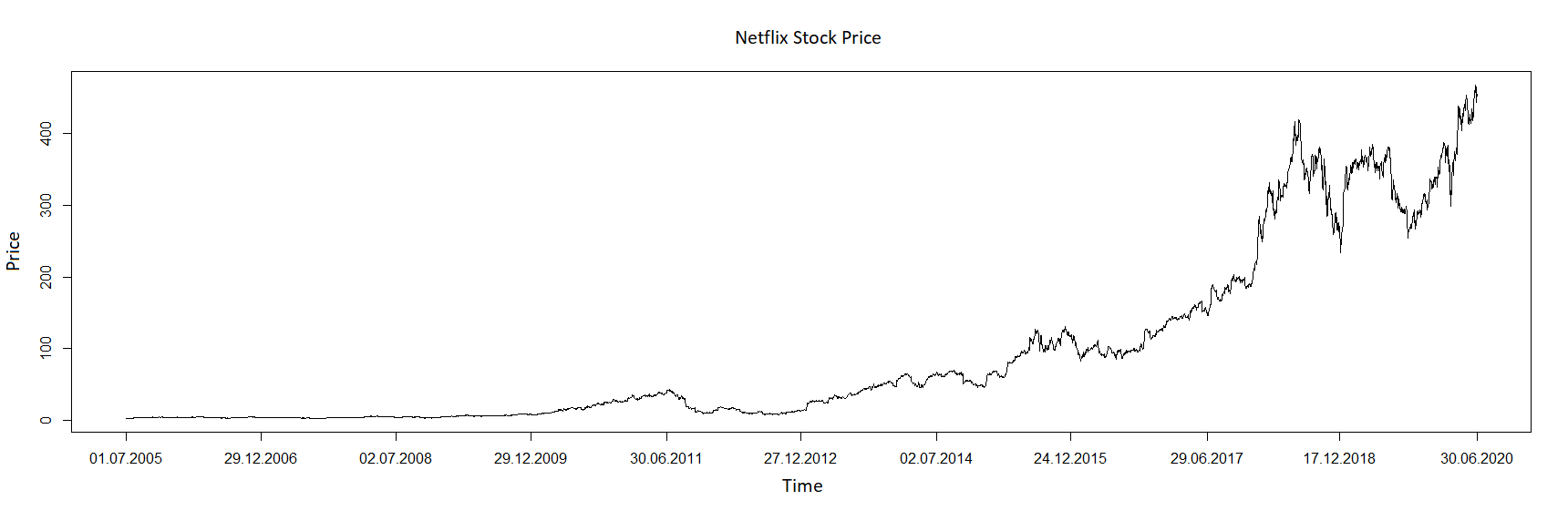}}
\subcaptionbox{Amazon Stock Price (USD) from July 1st 2005 to June 30 2020.}{\includegraphics[width=0.9\textwidth]{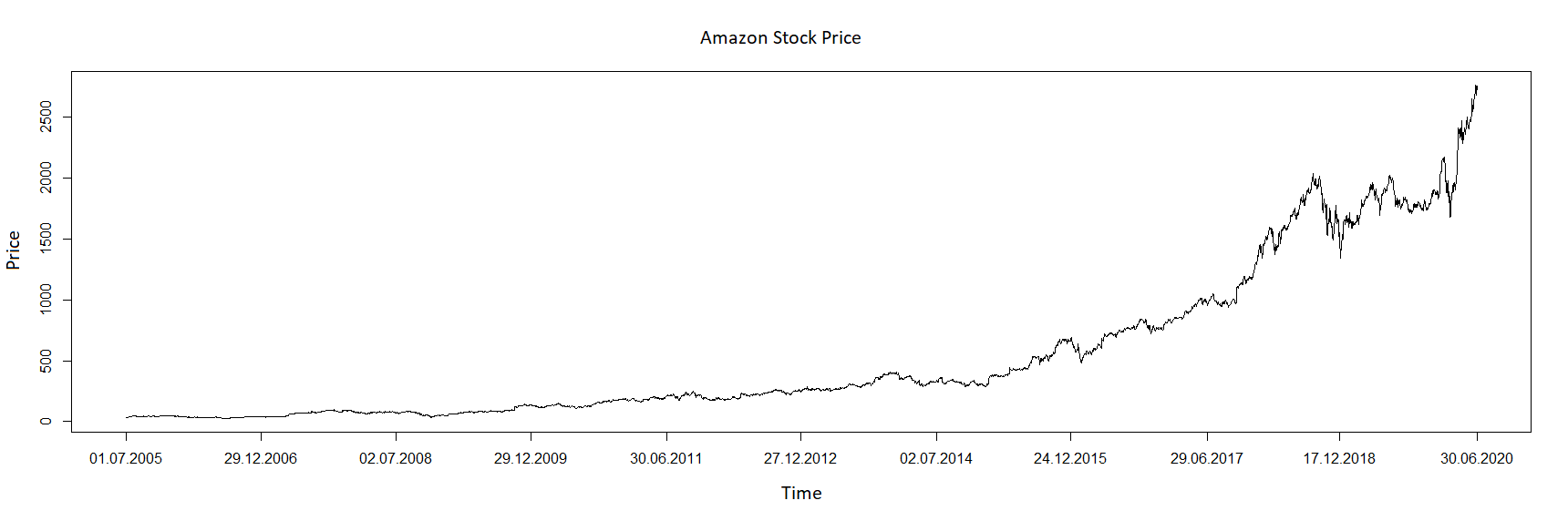}}
\caption{Amazon and Netflix stock prices, 15 year comparison.}
\label{fig:ComparacNFLXAMZN}
\end{figure}

In 1976, Merton \cite{Merton76} proposed a JDM for asset pricing to incorporate rare and abrupt changes in the values of $\{S_t\}_{t\ge 0}$. The JDM adds, to the GBM, a compound Poisson process where jumps are independent and identically distributed; see equation \eqref{EDJP} for a detailed definition. In Merton's model the logarithm of a jump is normally distributed \cite{Merton76}, which renders some computations relatively easier. On the other hand, Kou \cite{Kou03} modifies Merton's JDM by replacing the normal distribution with the Laplace distribution, this becomes computational more stable since the latter distribution is determined by one parameter only. Nevertheless, the continuous term of these models have constant drift and volatility which does not completely reflect the behavior of real data; see, e.g., Figure~\ref{fig:ComparacNFLXAMZN}.  

\begin{figure}
    \centering
    \includegraphics[width=0.7\textwidth]{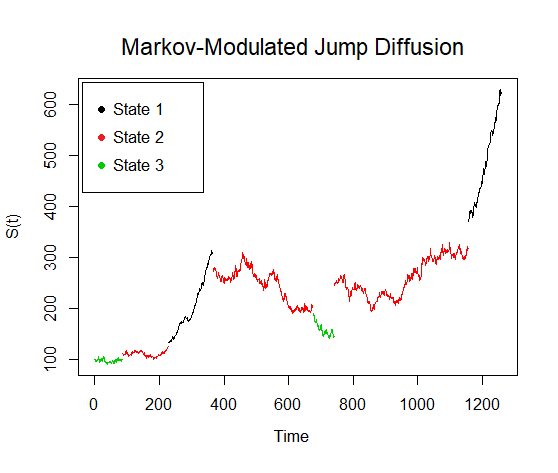}
    \caption{Path sample of a MMJDM; see Section~\ref{sec:Simulation Study} for details on the parameters. } 
    \label{fig:TrayecRepresent}
\end{figure}

Later models introduce random drift and random volatility parameters. e.g., in \cite{Guo01} and \cite{Job06}, the authors introduced the MMJDM which seeks to incorporate more variables that may influence the values of the stock prices; see Figure~\ref{fig:TrayecRepresent} for an example. In the MMJDM there is an underlying MJP that modulates both the drift and volatility of $\{S_t\}_{t\ge 0}$. The heuristic for this mechanism is that, abrupt jumps in the prices correspond to changes in the macroeconomic state of the market, therefore, not only a discontinuity occurs but also the process enters a distinct phase that determines distinct drift and volatility. We assume that the state space of the MJP represents the different environments of the financial market and so it influences the drift and volatility of the diffusion. MMJDMs for asset dynamics have been investigated in many works including \cite{Cos14}, \cite{Das18} and \cite{Ell07}.

In this paper, we are concerned with the estimation of parameters of a MMJDM. Inference, when there are two states in the underlying MJP, was presented in \cite{Das21}. Likelihood based estimation (including Bayesian) for MMJDM in risk theory has been investigated by \cite{Bal19} and \cite{Bal22}. Our main contribution is Section \ref{sec:Algorithms}, where we present a methodology for the inference of the MMJDM for the general case with $m\in\mathbb{N}$ states and Laplace distribution for the logarithm of each jump; see Section~\ref{sec:Model} for a detailed definition. We deal with maximum likelihood estimation in the situation where we observe the diffusion process (at discrete times) and no direct observations of the MJP are given. 
Therefore, the data can be viewed as incomplete observations from a model with a tractable likelihood function where the full data set consists of a continuous time record of both the diffusion process and the MJP. We therefore use the EM-algorithm (see, e.g. \cite{Mcl97}, \cite{Tanner96}) to tackle the difficulties that arise from using incomplete information in finding the MLEs. Another natural methodology used to estimate the parameters is through Markov chain Monte Carlo (MCMC) algorithms (see, e.g, \cite{Bal19}). Although the MCMC method allows us to obtain more information on the parameters from their distribution, in general, EM algorithms have two advantages: the speed of convergence and precision of the parameters are better and the estimation does not depend on the choice of the prior distribution.

In the next section we define the MMJDM. The methodology to estimate the parameters of the model is detailed in Section \ref{sec:Algorithms}. Sections \ref{sec:Simulation Study}~and~\ref{sec:Real Data} contain the simulation study and an analysis of real data from Amazon and Netflix, respectively; we close with several conclusions in Section \ref{sec:Conclusions}. 

\section{Markov-Modulated Jump- Diffusion Model}\label{sec:Model}
Before introducing the MMJDM, we present the definitions of both the GBM and the JDM. The GBM is the stochastic process $ \{S_t\}_{t \geq 0} $ solution of the stochastic differential equation (SDE)
\begin{equation}\label{EGBM}
dS_t=\mu S_tdt + \sigma S_t dB_t,
\end{equation} 
where $S_0>0$, $\mu\in\mathbb{R}$, $\sigma>0$ and $\{B_t\}_{t \geq 0}$ is a standard Brownian motion.

The process $\{S_t\}_{t \geq 0}$ is a JDM if it is given by the SDE
\begin{equation}\label{EDJP}
    dS_t =\mu S_t dt +\sigma S_t dB_t + d\sum_{k=1}^{N_t} \Delta S_k,
\end{equation}
where, in addition to the initial value $S_0$, parameters $\mu,\sigma$ and $\{B_t\}_{t \geq 0}$ we also have a compound Poisson process $\sum_{k=1}^{N_t} \Delta S_k$ with intensity $\lambda>0$ and i.i.d.~jump sizes $\Delta S_k$. We recall, as a warm up, that the solution for the JDM \eqref{EDJP} is
\begin{equation*}
S_t=S_0\exp\left(\left(\mu-\frac{1}{2}\sigma^2\right)t+\sigma B_t+\sum_{k=1}^{N_t} \Delta S_k\right).
\end{equation*}


For the MMJDM, we assume that, for some positive integer $m$, the financial market undergoes a macroeconomic environment that evolves among $m$ distinct states. The financial environment is then given by a MJP $\{X_t\}_{t\ge 0}$ with state space $E:=\{1,\ldots, m\}$ and intensity matrix $Q:=\{q_{ij}\}_{i,j\in E}$. Let $\{\tau_k\}_{k\ge 1}$ and $\{C_t\}_{t\geq 0}$ be, respectively, the jump times and the counting process that records the number of jumps of $\{X_t\}_{t\geq 0}$.

Each of the jumps of the MJP represents the occurrence of abrupt events that cause changes in the environment and we assume that state $i\in E$ represents the $i$-th most favourable and stable environment; from good growth and confidence to crisis in the business sector.

For the stock prices $\{S_t\}_{t\ge 0}$, we assume to have an initial state $S_0>0$ and vectors $\bm\mu:=(\mu_1,\ldots,\mu_m)$ and $\bm\sigma:=(\sigma_1,\dots,\sigma_m)$ corresponding to the drift and volatility of the stock prices at each state $i\in E$. In addition, let $\{J_k\}_{k\in\mathbb{N}}$ be a succession of independent random variables with Laplace distribution $(0,\eta)$; these variables will determine the jumps of the stock prices. The MMJDM associated to the stock prices is the solution to the SDE given by  
\begin{equation}\label{EOM}
dS_t = \mu_{X_t} S_t dt + \sigma_{X_t} S_tdB_t + d\sum_{k=1}^{C_t} \Delta S_k
\end{equation}
where $\Delta S_k:=\Big(\exp(J_k)-1\Big)S_{\tau_k^-}$, and $\{B_t\}_{t\geq0}$ is independent of $\{X_t\}_{t\geq 0}$. Note that the parameters of the MMDJM are given by $\bm \theta:=(Q,\bm\mu,\bm\sigma,\eta)$.

\section{Methodology for the Maximum Likelihood Estimators}\label{sec:Algorithms}

We deal with the situation where we do not have full observation of the MMJDM. The data are observations $S_{t_i}$ of the stock prices at discrete times $0=t_0<t_1\ldots<t_M=:T$, where $\Delta:=t_i-t_{i-1}$ for $i\in\{1,\ldots,M\}$. Our aim is to find the MLE of $\bm \theta$ when we only observe $\bm S:=\{S_{t_i}\}_{i=0}^M$; we consider this data as an incomplete observation of a full data set consisting of both sample paths $\bm S^c:=\{S_t\}_{t\in[0,T]}$ and $\bm X^c:=\{X_t\}_{t\in[0,T]}$.

First, we find the likelihood function for the full data set and then estimate conditional expectation of this full log-likelihood function given the observations $\bm S$. Assume that we count with $(\bm X^c,\bm S^c)$; that is, a continuous observation of the MJP and historical data of the stock prices. Then the likelihood function is given by 
{\small
\begin{equation*}
L_{T}^c(\bm \theta) := \mathbb{P}(\bm S^c,\bm X^c;\bm\theta)=\mathbb{P}(\bm S^c|\bm X^c;\bm\theta)\mathbb{P}(\bm X^c;\bm\theta);
\end{equation*}}
which may be expressed as
{\small{\small
\begin{equation}\label{lik_com}
 L_{T}^c(\bm \theta)=  \prod_{i=1}^m\left[\prod_{j=1}^{K}\eta e^{-\eta \ell_{j}}\left(\prod_{j=1}^{N_i}\prod_{k=2}^{r_{ij}}\frac{\exp\left(-\frac{(y_{ijk}-\delta_j\Delta)^2}{2\Delta\sigma_j^2}\right)}{\sqrt{2\pi\Delta\sigma_j^2}}\left(\prod_{j\neq i}^Mq_{ij}^{N_{ij}}e^{-q_{ij}R_i}\right)\right)\right],
\end{equation}
}}
where
\begin{itemize}
\item $N_{ij}$ is the number of jumps from state $i$ to $j$,
\item $N_i$ is the number of visits to state $i$, i.e., $N_i:=\sum_{j=1}^mN_{ij},$
\item $R_i$ is the total time spent at state $i$,
\item $K$ is the number of jumps $K:=\sum_{i=1}^mN_i$,
\item $r_{ij}$ is the number of observations of $\bm S$ on the $j$-th visit of $\bm X$ at state $i$, 
\item $y_{ijk}:=\mbox{log}\big(\frac{s_{ijk}}{s_{ij(k-1)}}\big)$ where $s_{ijk}$ is the $k$-th observation of stock prices on the $j$-th visit of $\bm X$ at state $i$,
\item $\ell_j$ is the size of the $j$-th jump, 
\item $\delta_j:=\mu_j-\frac{1}{2}\sigma_j^2$.
 \end{itemize}

In the case where the MMJDM is continuously observed, which is not a realistic assumption, finding the MLE is straightforward. For the case of incomplete data, the overall methodology is summarize in Algorithm~\ref{ALG-GEN}; we detail each of its steps in the subsequent sections.

\begin{algorithm}[H]
  \begin{algorithmic}[1]
\STATE Select a threshold to identify the jumps of the MJP. 
\STATE Estimate the coefficient of the jumps, i.e., compute the MLE of $\eta$.
\STATE Estimate the coefficients of the GBM, i.e., compute the MLE of $\bm\mu$ and $\bm\sigma$.
\STATE Estimate the intensity matrix of the MJP, i.e., compute the MLE of $Q$. 
\end{algorithmic}
  \caption{General algorithm when there are $m$ macroeconomic states\textbf{}}\label{ALG-GEN}
\end{algorithm}

\subsection{Identify the jumps of the MJP}
 
We will transfer the analysis of $\bm S$ to the log-yields $\bm Y:=\{y_1,\ldots,y_M\}$ which are defined by $y_i=\mbox{log}\big(\frac{S_{t_i}}{S_{t_{i-1}}}\big)$, for  $i=1,\ldots, M$.

Observe that if there are no jumps in  $(t_{i-1},t_i]$ we may write $X_t=j$ for $t\in (t_{i-1},t_i]$ and some $j\in E$ and so the process $\{S_t\}_{t\in(t_{i-1},t_i]}$ may be regarded as a GBM process with parameters $\mu_{j}$ and $\sigma_{j}$; see \eqref{EGBM}. 

Thus, conditional on not having jumps in $(t_{i-1},t_i]$ and using that $\{B_t\}_{t\geq 0}$ and $\{X_t\}_{t\geq 0}$ are independent, we can solve the SDE in \eqref{EOM} and obtain, for $t\in (t_{i-1},t_i]$,
\begin{equation}\label{MMJDsol}
    S_t=S_{t_{i-1}}\exp\left[\delta_{j} (t-t_{i-1})+\sigma_{j} (B_t-B_{t_{i-1}})\right].
\end{equation}
Consequently, for $i=1,\ldots,M$, assuming that $X_{t}=j\in E$ for $t\in[t_{i-1},t_i]$ (i.e., there are no jumps in such interval), we infer that $y_i\sim N(\delta_j\Delta,\sigma_j^2\Delta)$; that is, 
\begin{equation}\label{eq:logResuelto}
    y_i=\delta_j\Delta+\sigma_j (B_{t_i}-B_{t_{i-1}}).
\end{equation}

Heuristically, the logarithm of each jump $J_k$ should be significantly larger than the fluctuations of the diffusion over intervals of length $\Delta$, at any of the MJP states. Namely, we assume that $\eta^{-1}$ is significantly larger than $\max\{|\delta_j|:j\in E\}$.  
Consequently, we choose a suitable constant $U>0$ and define $\bm Y_J:=\{y_i\in Y:\, |y_i|>U\}$; we determine that $\bm S^c$ has (at least) a jump in the time interval $(t_{i-1},t_i]$ whenever $|y_i|>U$. 

\subsection{Estimate the distribution of the jumps}\label{sec:eta}

Write $\bm Y_J:=\{y_{i_1},...,y_{i_K}\}$. By assuming that there is at most one jump in each of the intervals $(t_i-t_{i-1}]$, we may associate $y_{i_k}$ to the $k$-th jump of the process; in fact, we heuristically argue below that $y_{i_k}\approx J_k$. Since $|J_k|\sim Exponential(\eta)$ we have that the MLE of  $\eta$ is given by
\begin{equation}\label{eq:EMVSaltoDif}
    \hat{\eta}:=\frac{K}{\sum_{k=1}^{K} |y_{i_k}|}.
\end{equation}
Indeed, assume that the $k$-th jump is contained in $(t_{i-1},t_i]$. We may assume that $X_{t_{i-1}}=j$, $X_{t_i}=j'$ for some distinct $j,j'\in E$. Hence, $y_i\in \bm Y_J$ is the sum of $J_k$ and two perturbations; that is 
\begin{equation*}
    y_i = J_k +\Delta P_{i-1,j,k}+\Delta P_{i,j',k};
\end{equation*}
where $\Delta P_{\ell,m,k}\sim N\left( \delta_m |t_\ell- \tau_k|, \sigma_m^2|t_\ell-\tau_k|\right)$. 
Note that $J_k$ is independent of both $\Delta P_{i-1,j,k}$ and $\Delta P_{i,j',k}$. The variability of the distribution of $J_k$ is assumed to be significantly larger than 
the fluctuations of the continuous part of the process in an interval of length $\Delta$. We have that 
\begin{equation*}
\left|\frac{\Delta P_{n,1}+\Delta P_{n,2}}{J_n}\right|\approx 0
\end{equation*}
which implies that the perturbations may be ignored, equivalently, $y_i \approx J_k$. 
 
\subsection{Estimate the coefficients of the GBM}

Let us analyse now the yields that correspond to intervals with no jumps. If $y_i\in \bm Y\setminus \bm Y_J$ then $X_t=j$ for some $j\in E$ and $t\in (t_{i-1},t_i]$. Therefore, 
\eqref{eq:logResuelto} implies that $y_i\sim N(\delta_{j}\Delta,\sigma_{j}^2\Delta)$.
Hence, we may regard $y_i$ as an observation of a mixture of normal distributions, and so we estimate the parameters $\bm \mu$ and $\bm \sigma$ from the observations $\bm Y\setminus \bm Y_J$.

For ease of writing, let $N:=M-K$ and let $\bm{W}_N:=\bm Y\setminus\bm Y_J=\{w_1,\ldots,w_N\}$ be an enumeration of the yields that preserves the original order in $\bm Y$. In what follows we introduce a latent discrete variable $\bm Z:=\{z_1,\ldots, z_N\}$ such that $z_n=j$ whenever $w_n\sim N(\delta_{j},\sigma_{j}^2)$; in this case we say that $w_n$ belong to group $G_j$. Observe that, conditioned to having no jumps in the interval corresponding to $w_n$, the latent variable $z_n$ coincides with $X_t$, for all $t$ in such interval.

For classifying $\bm{W_N}$, we can view each element $w_n$, with $n= 1,\ldots, N$, as an observation of a random variable $W$ with a $m$-component finite mixture density given by

\begin{equation}\label{eq:MezclaSinObs}
    f_W(w;\Psi):=\sum_{j=1}^m \pi_j \frac{1}{\sqrt{2\pi\sigma_j^2\Delta}}\exp\left(-\frac{1}{2\sigma_j^2\Delta}(w-\delta_j\Delta)^2\right),
\end{equation}
where $\Psi:=(\pi,\bm\mu,\bm\sigma)$ and $\pi_j$ are the mixing proportions; that is, $\sum_{j=1}^m \pi_j =1$ and for $j\in E$, $\mathbb{P}(Z_n=j)=:\pi_j\geq0$. We have that

\begin{equation}\label{eq:Gammas}
    \gamma_{Z_n}(j):=\mathbb{P}(Z_n=j|w_n)=\frac{\pi_jf_j(w_n;\psi_j)}{\sum_{k=1}^m \pi_k f_k(w_n;\psi_k)}, 
\end{equation}
where $\psi_j=(\mu_j,\sigma_j)$ and $f_j(w_n;\psi_j)$ is the probability density function of a normal distribution of parameters $(\delta_j\Delta,\sigma^2_j\Delta)$.

We shall call $({\bm W_N,\bm Z })$ the complete data set and refer to the actual observed data $\bm W_N$ as incomplete. We apply the EM algorithm to classify $\bm W_N$ and, simultaneously, approximate the MLE of both $\bm\mu$ and $\bm\sigma$. We have that the log-likelihood of the complete data (see \cite{GuzmanC18}) is given by
\begin{equation}\label{eq:EspLogVerosimMezcla}
\ell(\Psi|\bm W_N,\bm Z)=\ell(\Psi)=\sum_{n=1}^N\sum_{j=1}^m\log\left(\pi_j \gamma_{Z_n}(j);\psi_j)\right)+\sum_{n=1}^N\sum_{j=1}^m\log\left(f_j(w_n|Z_n=j;\psi_j)\right).
\end{equation}

Since the increments of the Wiener process $\{B_t\}_{t\geq0}$ are conditionally independent (on MJP), so are their corresponding log-returns 
and we can use the EM Algorithm \ref{alg:EMMezcla} to estimate $\bm\mu$ and $\bm\sigma$.

\begin{algorithm}[H]
  \begin{algorithmic}[1]
\STATE  Choose an initial value  $\Psi^{(0)}:=(\pi_j^{(0)},\delta_j^{(0)},{\sigma_j^2}^{(0)})_{j=1,...m}$, tolerance $\varepsilon >0$ and set $t=0$.
   
    \STATE{\bf E-step}: Compute $S(\Psi;\Psi^{(t)}):=\mathbb{E}[\ell(\Psi)|\bm W_N;{\Psi^{(t)}}]$ using \eqref{eq:EspLogVerosimMezcla} and set $$\gamma_{z_n}(j)^{(t+1)}=\displaystyle{\frac{\pi_j^{(t)} f_j(w_n;\psi_j^{(t)})}{\sum_{k=1}^m \pi_k^{(t)} f_k(w_n;\psi_k^{(t)})}},$$
    for $n=1,\ldots,N$ and $j\in E$.
    \STATE {\bf M-step}: For $j\in E$ let
    \begin{center}
        $\displaystyle{\pi_j^{(t+1)}=\frac{\sum_{n=1}^N\gamma_{z_n}(j)^{(t+1)}}{N},\hspace{0.25cm}\delta_j^{(t+1)}=\frac{\sum_{n=1}^N w_n\gamma_{z_n}(j)^{(t+1)}}{\sum_{n=1}^N\gamma_{z_n}(j)^{(t+1)}}},$\\
        ~\\$\displaystyle{{\sigma_j^2}^{(t+1)}=\frac{\sum_{n=1}^N (w_n-\delta_j^{(t+1)})^2 \gamma_{z_n}(j)^{(t+1)}}{\sum_{n=1}^N\gamma_{z_n}(j)^{(t+1)}}}.$
    \end{center}
    \STATE Compute $S(\Psi ;\Psi^{(t+1)})$ using \eqref{eq:EspLogVerosimMezcla}.
    \STATE If $|S(\Psi;\Psi^{(t)})-S(\Psi;\Psi^{(t+1)})|<\varepsilon$, let $\hat{\mu}_j=\delta_j^{(t+1)} +\frac{1}{2}{\sigma^2_j}^{(t+1)}$, $\hat{\sigma}_j=\sqrt{{\sigma_j^2}^{(t+1)}}$ and stop.
    \STATE Let $t=t+1$ and go to E-step.
  \end{algorithmic}\caption{ EM for estimation of  $\bm\mu$ and $\bm\sigma$}\label{alg:EMMezcla}
\end{algorithm}

\subsection{MLE of $Q$}

Finally, it remains to estimate the parameter $Q$ of the MJP process $\{X_t\}_{t\ge 0}$. To do so, we need to estimate the time the process spends at each state $j\in E$ and the frequency of jumps from state $i$ to state $j$, for distinct $i,j\in E$. First, we  classify  the clusters of yields between jumps, thus recovering the bulk of time spent at each state $j\in E$. The inference from the yields in which a jump is involved is more delicate and is performed in Algorithm~\ref{SEM_MJP}.

Let us decompose $\bm W_N$ into $K+1$ clusters, 
\begin{equation*}
    \{w_1^{0},\ldots,w_{n_{0}}^{0}\},\{w_1^{1},\ldots,w_{n_1}^1\},\ldots,\{w_1^{K},\ldots,w_{n_{K}}^{K}\},
\end{equation*}
where $\{w_1^k,\ldots,w_{n_k}^k\}$ are the yields between $k$-th and $(k+1)$-th jumps for $1\le k<K$ and, similarly,  $\{w_1^0,\ldots,w_{n_0}^0\}$ and $\{w_1^K,\ldots,w_{n_K}^K\}$ are the yields before the first jump and after the $K$-th jump, respectively.

We again refer to the latent variables $\bm Z$ to classify each cluster in one of the $m$ groups $G_j$, $j\in E$. For $j\in E$ and $0\le k\le K$, we have that the probability that the block $\{w_1^k,\ldots,w_{n_k}^k\}$ comes from the $j$-th density is
{\small
\begin{equation}\label{eq:ProbsClasificacion}
    \Gamma_k (j):=\mathbb{P}(z_1^k=j,\ldots,z_{n_k}^k=j|w_1^k,\ldots,w_{n_k}^k)=\prod_{\ell=1}^{n_k}\mathbb{P}(z_\ell^k=j|w_\ell^k)=\prod_{\ell=1}^{n_k} \gamma_{z_\ell^k}(j).
\end{equation}
}

Thus, we partially reconstruct a path of the MJP for the time intervals $\cup_{k=0}^K [r^-_k,r^+_k]$, where $r^-_k$ is the time of observation $w_1\textbf{}^k$ and  $r^+_k$ is the time of observation $w_{n_k}^k$. We classify each cluster $\{w_1^k,...,w_{n_k}^k\}$ using $\Gamma_k (j)$ for $j\in E$. If the $k$-th cluster is in state $j$ then $X_t=j$ for $t\in[r^-_k,r^+_k]$.

The values of $\{X_t\}$ are still undetermined in the missing intervals, i.e., we can only observe $\{X_t\}$ at the endpoints of interval $[r^+_k,r^-_{k+1}]$ for $k=0,\ldots,K-1$. Hence, we use Algorithm~\ref{SEM_MJP} to , iteratively, simulate such missing paths and estimate the MLE of each of the entries in the intensity matrix $Q$. Algorithm~\ref{SEM_MJP} is a Stochastic EM (SEM) algorithm (see, e.g. \cite{Nilsen00}); the E-step and M-step of each of these algorithms are repeated until convergence, i.e., when the estimators are observed to arrive at the stationary distribution. An advantage of using the stochastic version of EM algorithm is that, by having the distribution of the estimator, we can calculate confidence intervals for the MLE's obtained.

\begin{algorithm}[H]
  \begin{algorithmic}[1]
 \STATE Choose an initial value $\bm Q^0$ and make $\ell=0$.
 \STATE Using the current $\bm Q^\ell$, for each $0\le k<K-1$, draw a Markov bridge between the observations $X_{r^+_{k-1}}$ and $X_{r^-_{k}}$ to generate a path $\bm X^c=\{X_t\}_{t=0}^{T}$.
\STATE \textbf{E-step} 
Using $\bm X^c$, we compute the corresponding statistics $N_{ij}$ and $R_i$ for all $i,j\in E$ in time interval $[0,T]$.
\STATE \textbf{M-step} The MLEs of $q_{ij}$ (see \cite{Bal22}) are given by  $$q_{ij}^{\ell+1}= \frac{N_{ij}}{R_i},\hspace{.2cm}i\neq j,\hspace{.2cm} q_{ii}^{\ell+1}=-\sum_{j\neq i}q_{ij}^{\ell+1}.$$
\STATE  $\ell=\ell+1$ and go to Step 2. 
\end{algorithmic}
  \caption{SEM for the estimation of $\bm Q$.}\label{SEM_MJP}
\end{algorithm}

\section{Simulation Study}\label{sec:Simulation Study}

In this section, we present the results of a simulation study that calibrates the method presented in Section \ref{sec:Algorithms}.
We test the model with $m=3$; that is, we consider three states going from a stable economy to an  unstable economy to crisis in the sector. Hence, we may consider in the test that drifts $\mu_i$ are decreasing (possibly $\mu_m<0$) while volatilities $\sigma_i$ are increasing in $i\in E$.

We consider a MMJDM $\{S_t\}_{t\ge 0}$ with the following parameters. Let 
\begin{align}
\label{eq:GenEjemplo}
Q&=\begin{pmatrix} 
-0.002645503 & 0.002314815 & 0.00033068785\\
0.003968254 & -0.005952381 & 0.0019841270\\
0.005952381 & 0.001984127 & -0.0079365079\\
\end{pmatrix},\\
\bm\mu&=(0.0059523810, 0.0011904762, -0.0009920635), \label{eq:muEjemplo}\\
\bm\sigma&=(0.009449112, 0.017817415, 0.022047928), \label{eq:sigmaEjemplo}\\
\eta&=250/33. \label{eq:etaEjemplo}
\end{align}

We generate a sample path $\bm S= \{S_{t_i}\}_{i=0}^M$ where $S_0=100$, $X_0=1$, $\Delta=1$ and $T=1260,2520,3780,5040,6300,7560,8820$. Observe that if $\Delta=1$ represents one-day periods, then $\bm S$ represents, for example, the daily observations of the stock prices during a 7-year period of time if $T=8820$. In fact, the parameters $\bm\theta$ may be proposed on a time scale of 1 year (252 business days) and then rescaled to \eqref{eq:GenEjemplo}--\eqref{eq:etaEjemplo} which then may be interpreted as daily rates of the process.

Figure \ref{fig:TrayecRepresent} shows a path of MMJDM $\{S_t\}_{t\ge 0}$ with the aforementioned parameters in the time interval $[0,1260]$. The MLEs of $\eta$ are presented in Table \ref{tab:EstimacionEta}. The MLEs of $\bm\mu$ and $\bm\sigma$ for 1000 iterations are given in Table \ref{tab:EstimacionMedias} and, Table  \ref{tab:EstimacionVolatilidades}  respectively; Algorithm \ref{alg:EMMezcla} was run with $\epsilon= 0.05$. We tried several seeds $\bm\Psi^{(0)}, \bm Q^{(0)}$ for Algorithms \ref{alg:EMMezcla} and \ref{SEM_MJP}. In this study, no dependence was observed in the estimations except for the number of iterations, so we report the results using the seeds that best represent the model.

\begin{table}
\centering
\begin{tabular}{|c|c|c|}
\hline
\rowcolor[HTML]{ECF4FF}
 & Value & Error \\ \hline
$\eta$ & 7.575757 & \\ \hline
$\hat{\eta}^{1260}$ & 5.897666 & 1.678092\\ \hline 
$\hat{\eta}^{2520}$ & 6.241154 & 1.334603\\ \hline
$\hat{\eta}^{3780}$ & 7.756247 & 0.1804897\\ \hline
$\hat{\eta}^{5040}$ & 7.352301 & 0.2234567\\ \hline
$\hat{\eta}^{6300}$ & 7.144313 & 0.4314449\\ \hline
$\hat{\eta}^{7560}$ & 6.994469 & 0.5812881\\ \hline
$\hat{\eta}^{8820}$ & 7.175147 & 0.4006109\\ \hline
\end{tabular}
\caption{MLE  of $\eta$}
\label{tab:EstimacionEta}
\end{table}

\begin{table}
\centering
\begin{tabular}{|c|c|c|c|c|}
\hline
\rowcolor[HTML]{ECF4FF}
 & $j=1$ & $j=2$ & $j=3$ & Error\\ \hline
$\mu_j$ & 0.0059523810 & 0.0011904762 & -0.0009920635 & \\ \hline
$\hat{\mu_j}^{1260}$ & 0.0059469097 & 0.0027668870 & 0.0019743179 & 0.00001128\\ \hline
$\hat{\mu_j}^{2520}$ & 0.0066422130 & 0.0034594822 & 0.0025009336 & 0.00001782\\ \hline
$\hat{\mu_j}^{3780}$ & 0.0066102728 & 0.0032372128 & 0.0020031847 & 0.00001359\\ \hline
$\hat{\mu_j}^{5040}$ & 0.0063053813 & 0.0028402039 & 0.0020858060 & 0.00001231\\ \hline
$\hat{\mu_j}^{6300}$ & 0.0059171341 & 0.0031332369 & 0.0017120427 & 0.00001108\\ \hline
$\hat{\mu_j}^{7560}$ & 0.0060495312 & 0.0030009829 & 0.0013917915 & 0.00000897\\ \hline
$\hat{\mu_j}^{8820}$ & 0.0060614045 & 0.0029881755 & 0.0008520252 & 0.00000664\\ \hline
\end{tabular}
\caption{MLE of $\mu_j$.}
\label{tab:EstimacionMedias}
\end{table}

\begin{table}
\centering
\begin{tabular}{|c|c|c|c|c|}
\hline
\rowcolor[HTML]{ECF4FF}
 & $j=1$ & $j=2$ & $j=3$ & Error\\ \hline
$\sigma_j$ & 0.009449112 & 0.017817415& 0.022047928 & \\ \hline
$\hat{\sigma_j}^{1260}$ & 0.012769167 & 0.015921710 & 0.017703704 & 0.00003348\\ \hline
$\hat{\sigma_j}^{2520}$ & 0.009535770 & 0.015774837 & 0.017788880 & 0.00002231\\ \hline
$\hat{\sigma_j}^{3780}$ & 0.009327915 & 0.015878260 & 0.018081729 & 0.00001950\\ \hline
$\hat{\sigma_j}^{5040}$ & 0.009104289 & 0.015999326 & 0.017911584 & 0.00002053\\ \hline
$\hat{\sigma_j}^{6300}$ & 0.009154862 & 0.015454727 & 0.018728368 & 0.00001668\\ \hline
$\hat{\sigma_j}^{7560}$ & 0.009237009 & 0.014904877 & 0.014904877 & 0.00001700\\ \hline
$\hat{\sigma_j}^{8820}$ & 0.014904877 & 0.014492026 & 0.019310947 & 0.00001854\\ \hline
\end{tabular}
\caption{MLE of  $\sigma_j$}
\label{tab:EstimacionVolatilidades}
\end{table}

\begin{figure}
\centering
\subcaptionbox{$q_{12}$ estimation\label{EjEstimq12}}{\includegraphics[width=0.46\textwidth]{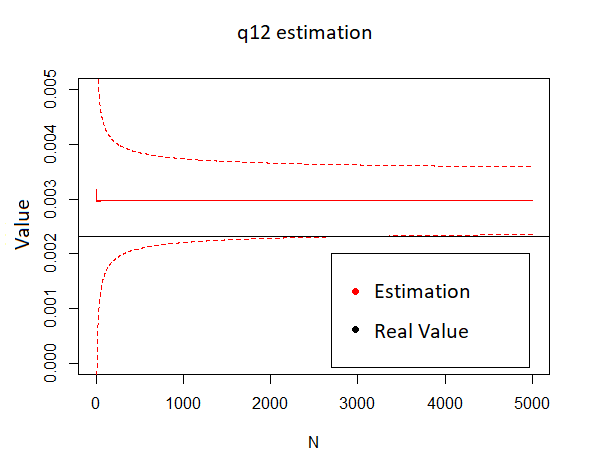}}
\subcaptionbox{$q_{13}$ estimation\label{EjEstimq13}}{\includegraphics[width=0.46\textwidth]{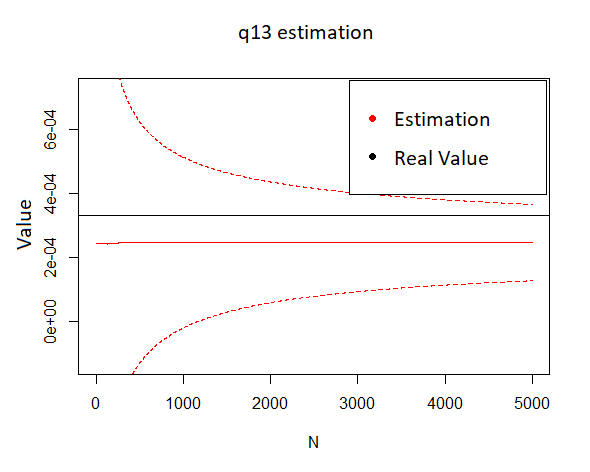}}
\subcaptionbox{$q_{21}$ estimation\label{EjEstimq21}}{\includegraphics[width=0.46\textwidth]{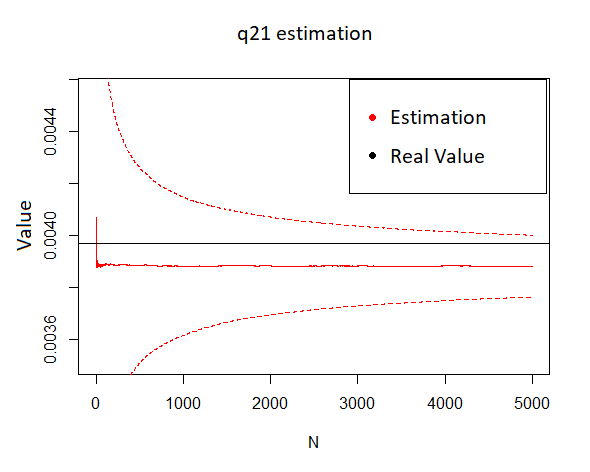}}
\subcaptionbox{$q_{23}$ estimation\label{EjEstimq23}}{\includegraphics[width=0.46\textwidth]{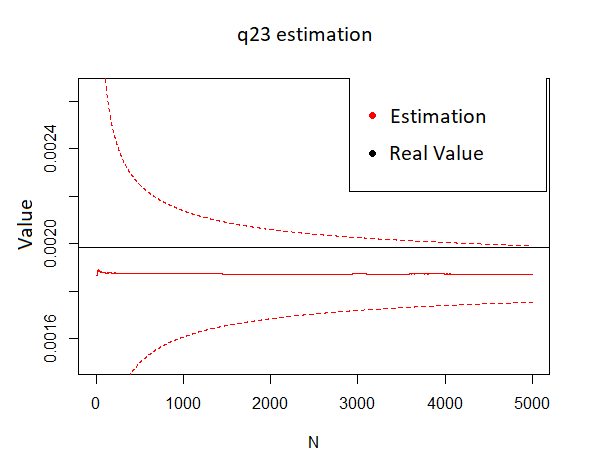}}
\subcaptionbox{$q_{31}$ estimation\label{EjEstimq31}}{\includegraphics[width=0.46\textwidth]{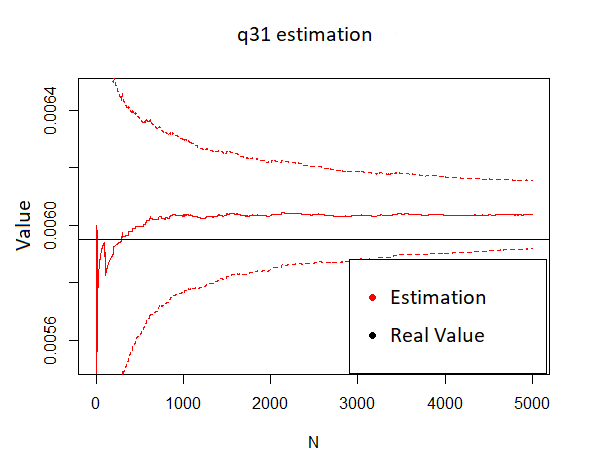}}
\subcaptionbox{$q_{32}$ estimation\label{EjEstimq32}}{\includegraphics[width=0.46\textwidth]{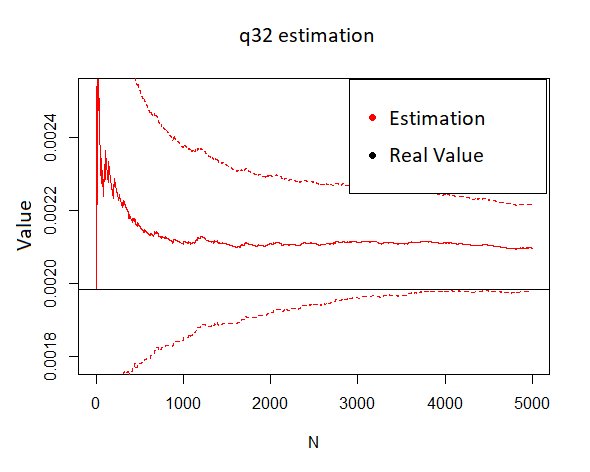}}
\caption{Estimation of $Q$ for the simulation study with $T=8820$ and 95$\%$ confidence intervals (doted line).}\label{fig:EstimGenEj}
\end{figure}

Finally, the MLE of the intensity matrix (after 1000 iterations) when $T=8820$ is given in equation \eqref{eq:EstimadorGenEjemplo}. Figure \ref{fig:EstimGenEj} plots the iterations of the Algorithm \ref{ALG-GEN} to find the MLE of $\hat{Q}$ and 95\% confidence intervals; we estimate the standard deviation of the MLEs through the Fisher information matrix (see \cite{Bla11}). We obtain

\begin{equation}\label{eq:EstimadorGenEjemplo}
\hat{Q}=\begin{pmatrix} 
-0.003215818 & 0.002969198 & 0.0002466196\\
0.003882034 & -0.005754555 & 0.0018725209\\
0.006037886 & 0.002097409 & -0.0081352957\\
\end{pmatrix}.
\end{equation}

\newpage
\section{Real Data}\label{sec:Real Data}
In this section, we fit the MMJDM for two sets of stock prices real data using the method presented in Section \ref{sec:Algorithms}. Each data set corresponds to Amazon and Netflix daily stock prices from July 1st, 2005 to June 30, 2020; see Figure~\ref{fig:ComparacNFLXAMZN}. 
Thus, it is natural to consider a day as the unit of time; that is $\Delta=1$. This historical data presents abrupt changes and these do not occur homogeneously in time and so we may use the MMJDM model to fit the data; based on the graphical evidence, we assume that the underlying MJP has three states.\footnote{ Estimating the number of states is a classification problem is more complex and beyond the scope of this paper.}

Following our method, the first step is to identify the jumps. We can observe in the final prices of Netflix's stock that the price suffers a change in its behavior after March 20, 2020. This can be explained because in March many countries introduced lockdown measures due to COVID 19 pandemic. In order to catch this change in this behaviour the threshold
$U$ for the jumps was set to $\pm \sqrt{0.01174792}$ for the squared log-returns. Similarly, the tolerance level for Amazon was set to $\pm\sqrt{0.005068828}$. As we do not consider jumps in consecutive days, we have 39 jumps for Netflix and 41 jumps for Amazon. These were used to obtain the jump parameter estimations $\hat{\eta}=6.359798$ for Netflix and $\hat{\eta}=9.119797$ for Amazon.

Algorithm \ref{alg:EMMezcla} was run considering 1000 iterations; we use $\epsilon=0.05$  and initial values as in \eqref{eq:muEjemplo} and \eqref{eq:sigmaEjemplo}. The MLEs obtained for the drift and volatility parameters for both Netflix and Amazon are shown in Table \ref{tab:ParamEstim}. 

\begin{table}
\begin{subtable}[t]{0.48\textwidth}
\begin{tabular}{|c|c|c|c|}
\hline
\rowcolor[HTML]{ECF4FF} 
j & $\hat{\pi}_j$ & $\hat{\mu}_j$ & $\hat{\sigma}_j$   \\ \hline
1 & $0.44797175$ & $-0.00026218670$ & $0.01402849$\\ \hline
2 & $0.53779080$ & $0.0031193879$ & $0.03213458$\\ \hline
3 & $0.01423745$ & $-0.00634970790$ & $0.09737263$\\ \hline
\end{tabular}
\caption{Netflix}
\label{tab:ParamEstimNFLX}
\end{subtable}
\hspace{\fill}
\begin{subtable}[t]{0.48\textwidth}
\flushright
\begin{tabular}{|c|c|c|c|}
\hline
\rowcolor[HTML]{ECF4FF} 
j & $\hat{\pi}_j$ & $\hat{\mu}_j$ & $\hat{\sigma}_j$   \\ \hline
1 & $0.45787160$ & $0.0009578106$ & $0.01093079$\\ \hline
2 & $0.48193159$ & $0.0012029475$ & $0.02213378$\\ \hline
3 & $0.06019681$ & $0.0041515128$ & $0.04834662$\\ \hline
\end{tabular}
\caption{Amazon}
\label{tab:ParamEstimAMZN}
\end{subtable}
\caption{\footnotesize MLE of the weights, drifts and volatilities for Netflix and Amazon obtained with Algorithm \ref{alg:EMMezcla} with tolerance level $\epsilon=0.05$.}
\label{tab:ParamEstim}
\end{table}

\begin{figure}[h!]
\centering
\subcaptionbox{$q_{12}$ estimation\label{NFLXEstimq12}}{\includegraphics[width=0.436\textwidth]{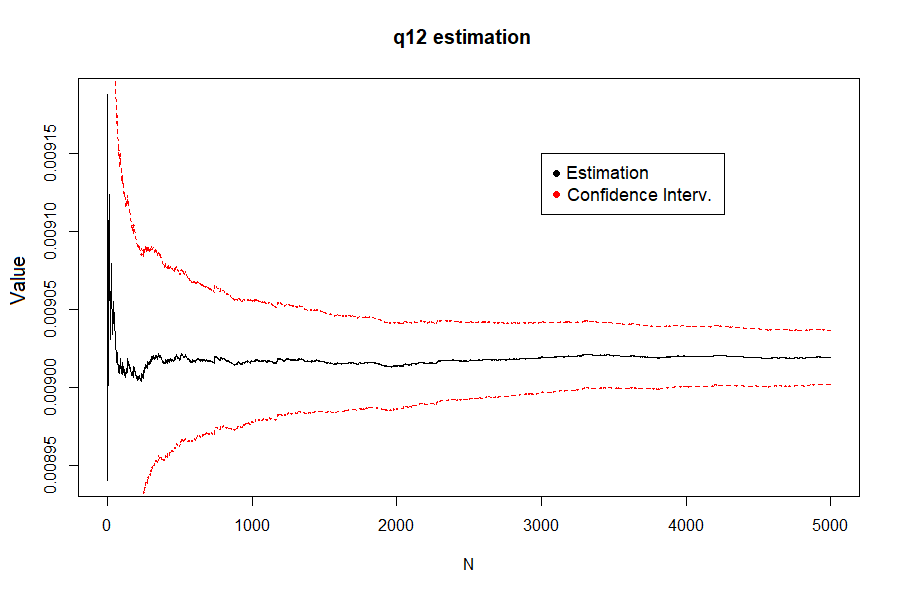}}
\subcaptionbox{$q_{21}$ estimation\label{NFLXEstimq13}}{\includegraphics[width=0.436\textwidth]{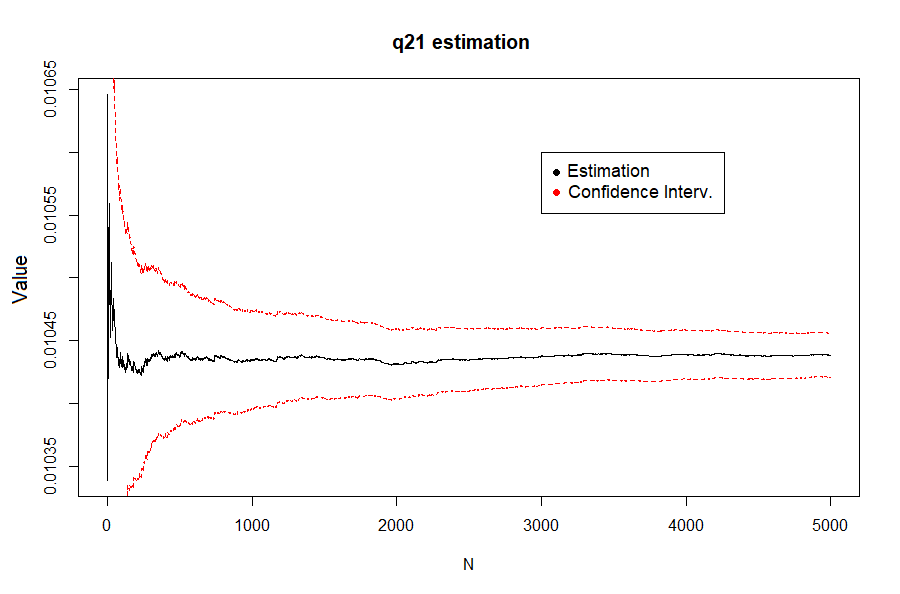}}
\subcaptionbox{$q_{23}$ estimation\label{NFLXEstimq21}}{\includegraphics[width=0.436\textwidth]{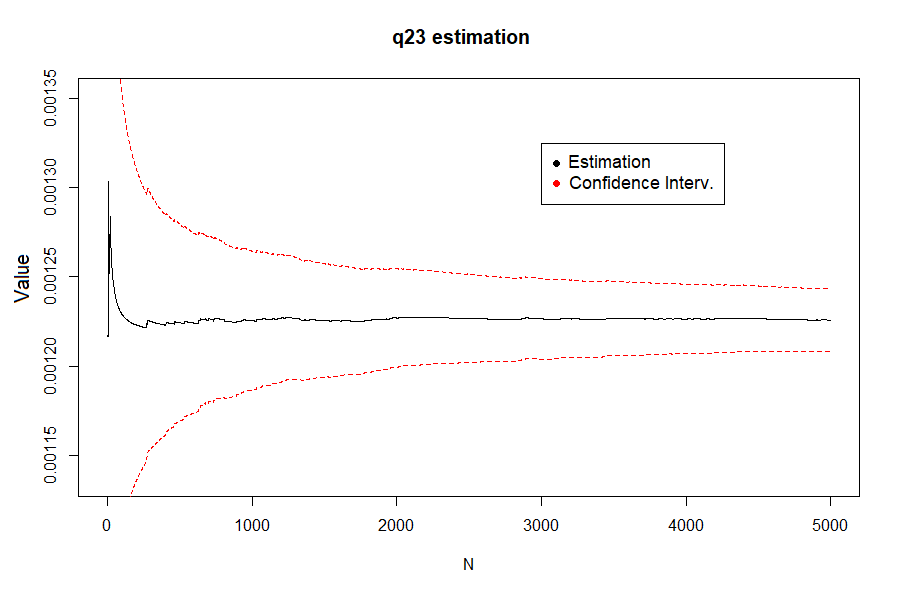}}
\subcaptionbox{$q_{32}$ estimation\label{NFLXEstimq23}}{\includegraphics[width=0.436\textwidth]{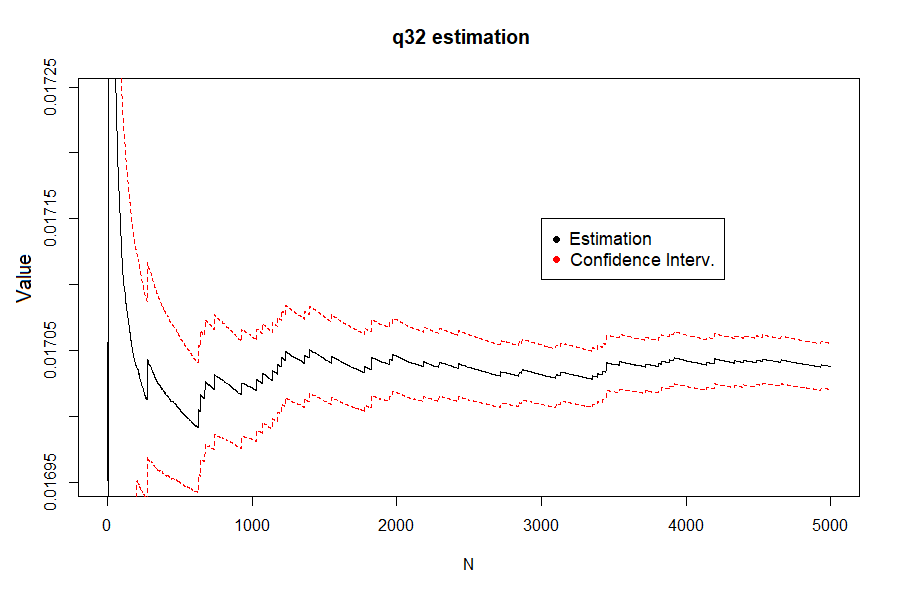}}
\caption{Estimation of $Q$ for Netflix stock.}\label{fig:EstimGenNFLX}
\end{figure}

\begin{figure}[h!]
\centering
\subcaptionbox{$q_{12}$ estimation\label{AMZNEstimq12}}{\includegraphics[width=0.436\textwidth]{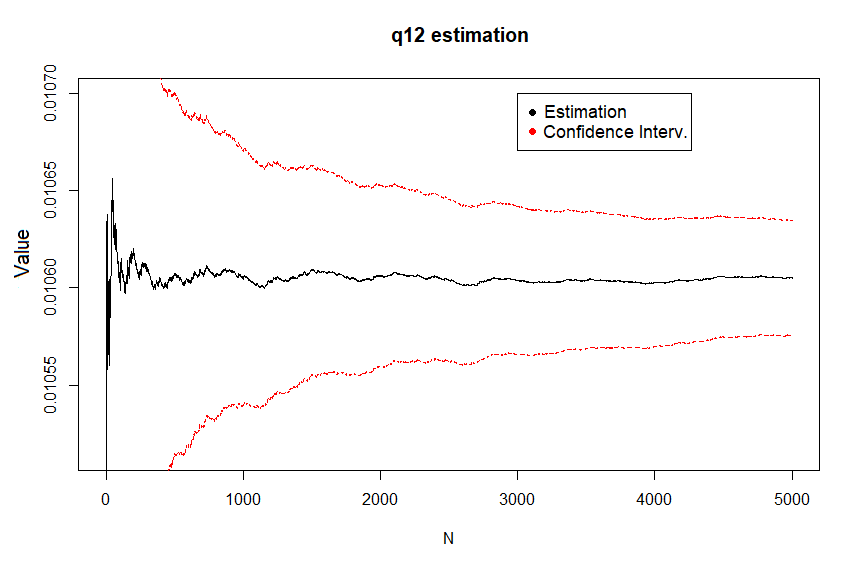}}
\subcaptionbox{$q_{21}$ estimation\label{AMZNEstimq13}}{\includegraphics[width=0.436\textwidth]{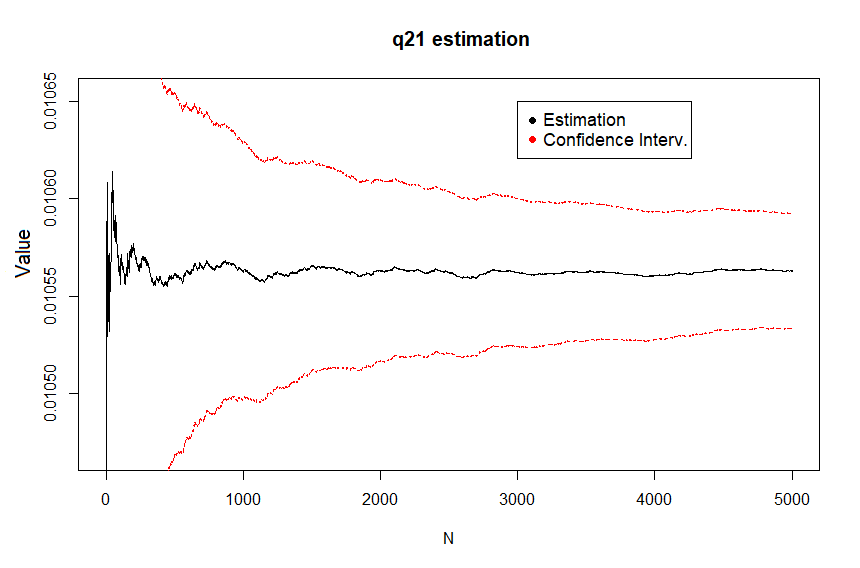}}
\subcaptionbox{$q_{23}$ estimation\label{AMZNEstimq21}}{\includegraphics[width=0.436\textwidth]{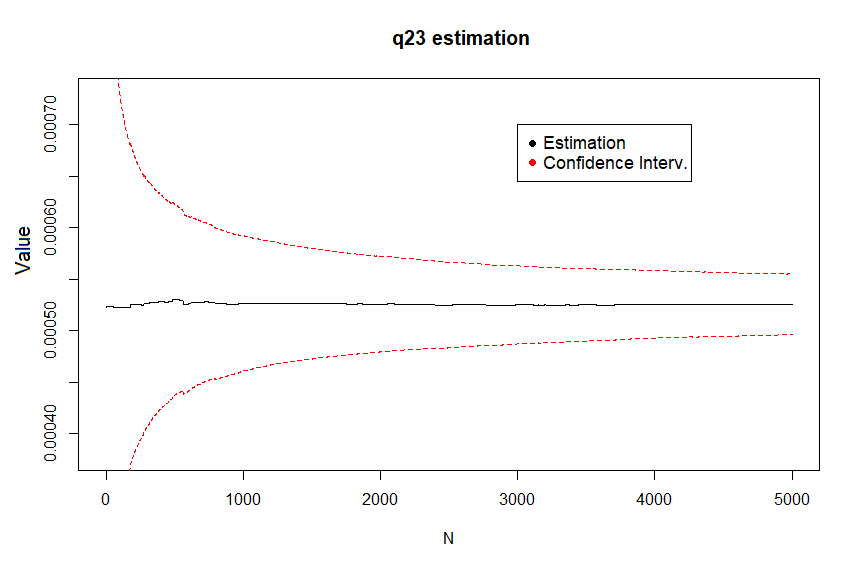}}
\subcaptionbox{$q_{32}$ estimation\label{AMZNEstimq23}}{\includegraphics[width=0.436\textwidth]{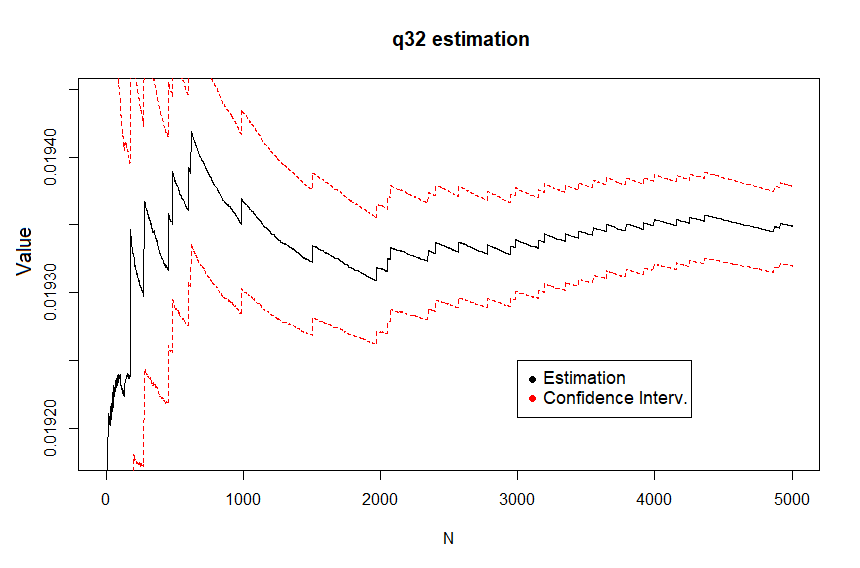}}
\caption{Estimation of $Q$ for Amazon stock.}\label{fig:EstimGenAMZN}
\end{figure}

Once the drifts and volatilities are estimated, we classified each cluster using \eqref{eq:ProbsClasificacion} and partially constructed the path of the MJP for each stock. 
The initial value for both $Q_{NFLX}$ and $Q_{AMZN}$ estimation was \eqref{eq:GenEjemplo} and we set the EM algorithm to 1000 initial iterations. The MLE for $Q_{NFLX}$ and $Q_{AMZN}$ is shown in equation \eqref{eq:GenNFLXComp} and \eqref{eq:GenAMZNComp}; respectively. Finally, the graphs of the estimations of $Q_{NFLX}$ and $Q_{AMZN}$ are shown in Figures \ref{fig:EstimGenNFLX} and \ref{fig:EstimGenAMZN}, respectively.



\begin{equation}\label{eq:GenNFLXComp}
\hat{Q}_{NFLX}=\begin{pmatrix} 
-0.009019071 & 0.009019071 & 0.000000000 \\
0.010438460 & -0.011664262 & 0.001225801\\
0.000000000 & 0.017037859 & -0.017037859\\
\end{pmatrix},
\end{equation}
\begin{equation}\label{eq:GenAMZNComp}
\hat{Q}_{AMZN}=\begin{pmatrix}
-0.01060526 & 0.01060526 & 0.0000000000\\
0.01056306 & -0.01108866 & 0.0005255957\\
0.00000000 & 0.01934920 & -0.0193491966\\
\end{pmatrix}.
\end{equation}


\section{Conclusions}\label{sec:Conclusions}

We propose a methodology based on Maximum Likelihood Estimators to estimate the parameters of the MMJDM process; this process models stock prices whose drift and volatility are shaped by macroeconomic states. 

There are two key properties of the stock market that the MMJDM describes. First, changes in the macroeconomic states may lead to discontinuities or jumps in the prices (disguised in the sample paths as abrupt fluctuations). Second, when these changes occur they also modify the overall diffusion properties of the prices. Although previous Jump-Diffusion Models incorporate the eventual jumps in the process, the MMJDM also takes into account the changes in drift and volatility that fortuitous events or news in the sector may entail. 

Among the challenges faced in this work are identifying the jumps, classifying the clusters of the log yields and optimizing the EM algorithms. We highlight that, as the simulation study in Section \ref{sec:Algorithms} shows, the methodology is robust since we are able to recover the parameters of the synthetic MMJDM. 

The study with the real data on both Amazon and Netflix' prices qualitatively reflects the same macroeconomic fluctuations. It is natural to conjecture that if the states are ordered according to the degree of stability in the market then the transitions between the stability and crisis will be sequential (e.g. before a crisis hit, there is a period of increased volatility). It is remarkable to note that, although this heuristic is not built into the model, there is no transition from the stable to the crisis state (that is; $\hat{q}_{13}=\hat{q}_{31}=0$).

Therefore we consider that the MMJDM more accurately represents the nature of fluctuations in the stock market. Moreover, the methodology proposed in this work to estimate the parameters of the process allows us to make inferences about the financial environments that favor a stock, which a useful tool in stock value speculation.

As future work, one may generalize this method by replacing the JMP to an underlying process where the length of stay in each stage follows a Phase Type distribution. One may also  fit the distribution of the jump sizes according to the observed data. Finally, one may also consider methodologies that test for a suitable number of macroeconomic states since, at the moment, we assume the number of states $m$ is given.

\section*{Acknowledgements}

We would like to thank the anonymous referee for their thorough  comments. This work has been supported by PAPIIT TA100820.


\end{document}